\documentclass[twocolumn,aps,showpacs,floatfix,superscriptaddress]{revtex4}
\usepackage{amsmath,amssymb,eucal,graphicx,subfigure}
\begin{document}

\title{Dynamics of an Idealized Model of Microtubule Growth and Catastrophe}
\author{T. Antal}
\affiliation{Program for Evolutionary Dynamics, Harvard University,
  Cambridge, MA 02138, USA}
\author{P. L. Krapivsky}
\affiliation{Center for Polymer Studies and Department of Physics, Boston University, Boston, MA 02215, USA}
\author{S. Redner}
\affiliation{Center for Polymer Studies and Department of Physics, Boston University, Boston, MA 02215, USA}
\author{M. Mailman}
\affiliation{Martin Fisher School of Physics, Brandeis University, Waltham, MA 02454, USA}
\author{B. Chakraborty}
\affiliation{Martin Fisher School of Physics, Brandeis University, Waltham, MA 02454, USA}
\begin{abstract}

  We investigate a simple dynamical model of a microtubule that evolves by
  attachment of guanosine triphosphate (GTP) tubulin to its end, irreversible
  conversion of GTP to guanosine diphosphate (GDP) tubulin by hydrolysis, and
  detachment of GDP at the end of a microtubule.  As a function of rates of
  these processes, the microtubule can grow steadily or its length can
  fluctuate wildly.  In the regime where detachment can be neglected, we find
  exact expressions for the tubule and GTP cap length distributions, as well
  as power-law length distributions of GTP and GDP islands.  In the opposite
  limit of instantaneous detachment, we find the time between catastrophes,
  where the microtubule shrinks to zero length, and determine the size
  distribution of avalanches (sequence of consecutive GDP detachment events).
  We obtain the phase diagram for general rates and verify our predictions by
  numerical simulations.

\end{abstract}
\pacs{87.16.Ka, 87.17.Aa, 02.50.Ey, 05.40.-a}

\maketitle

\section{INTRODUCTION AND MODEL}

\begin{figure}[ht]
\includegraphics*[width=0.39\textwidth]{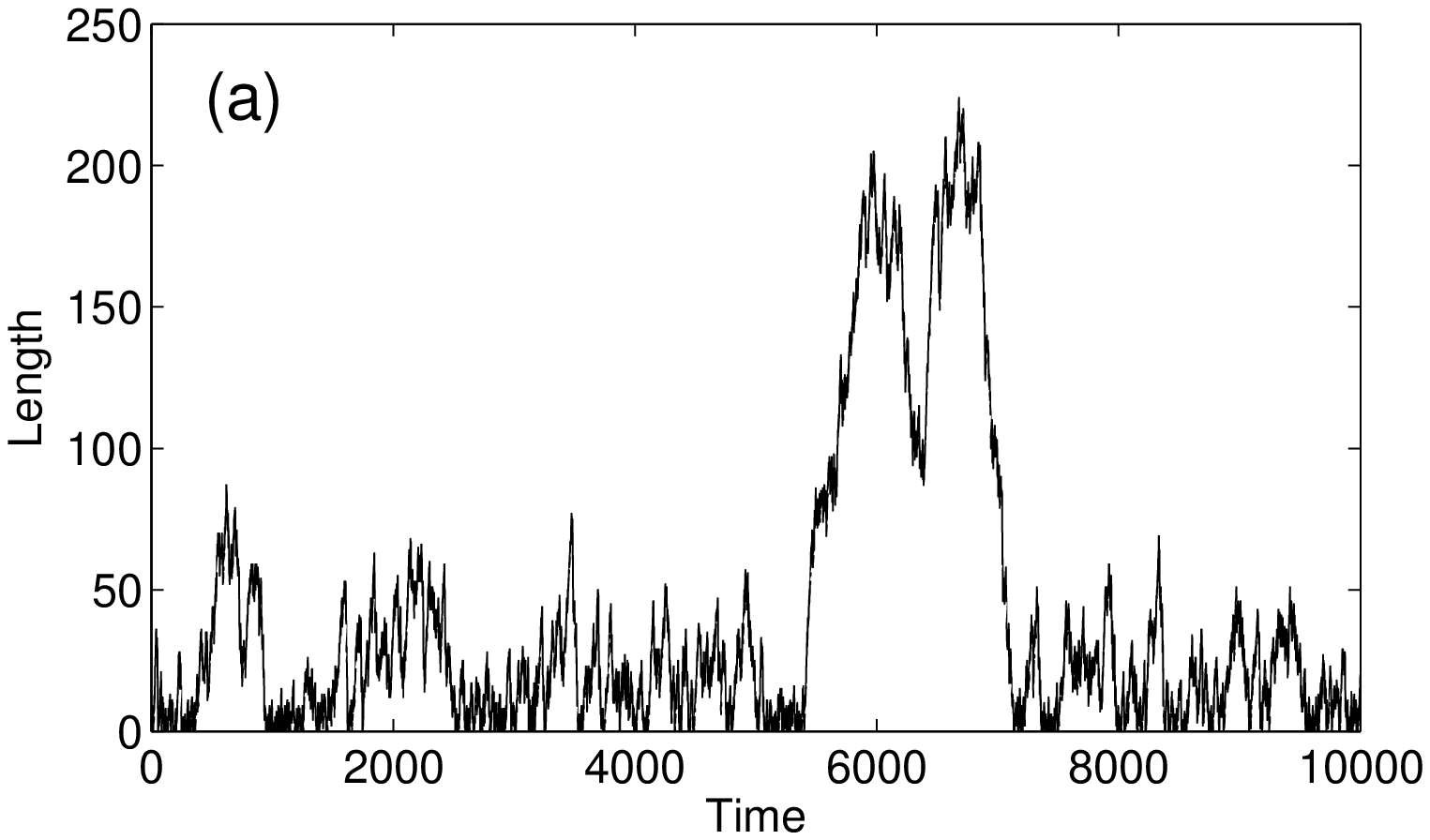}
\includegraphics*[width=0.39\textwidth]{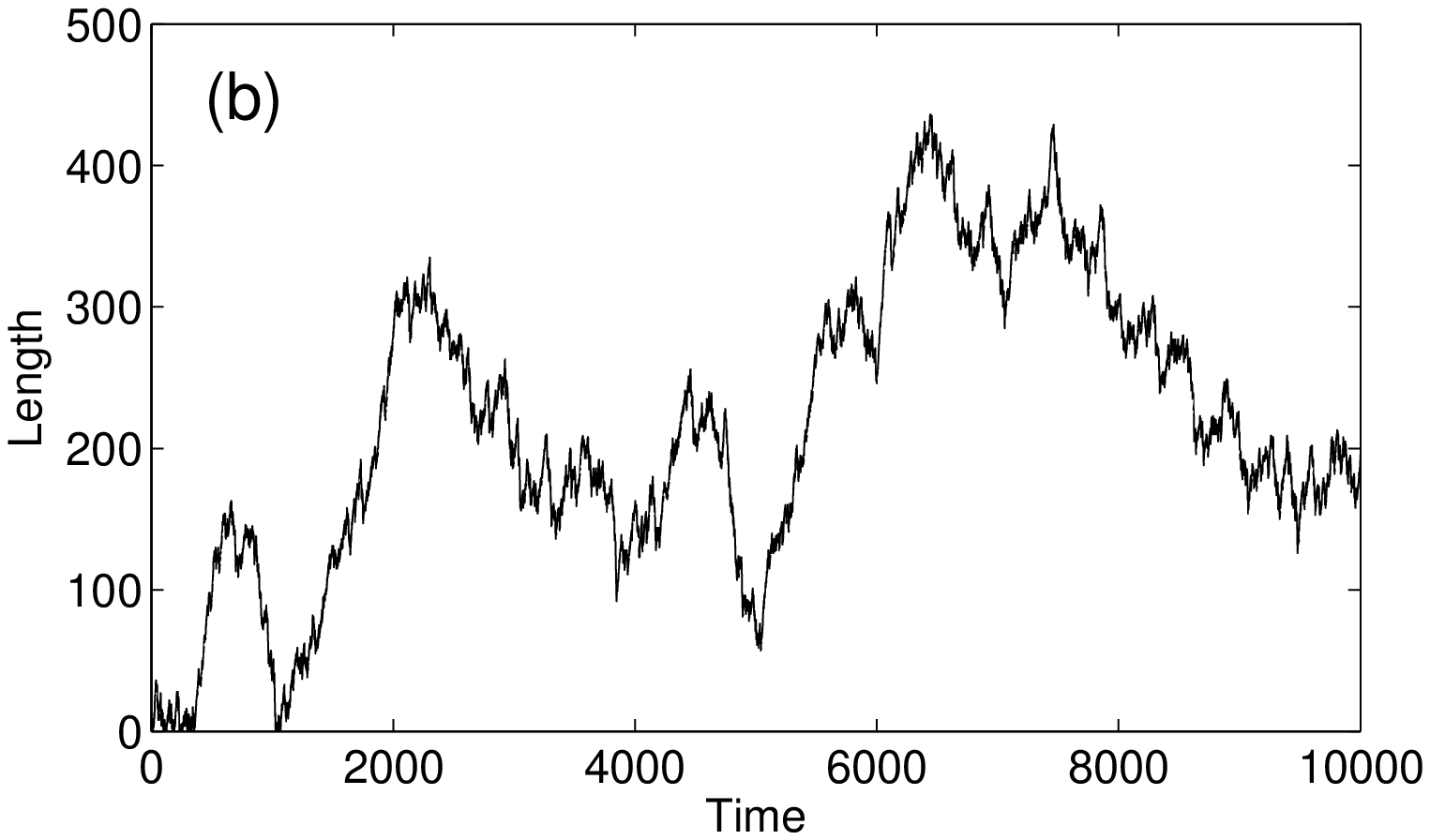}
\includegraphics*[width=0.39\textwidth]{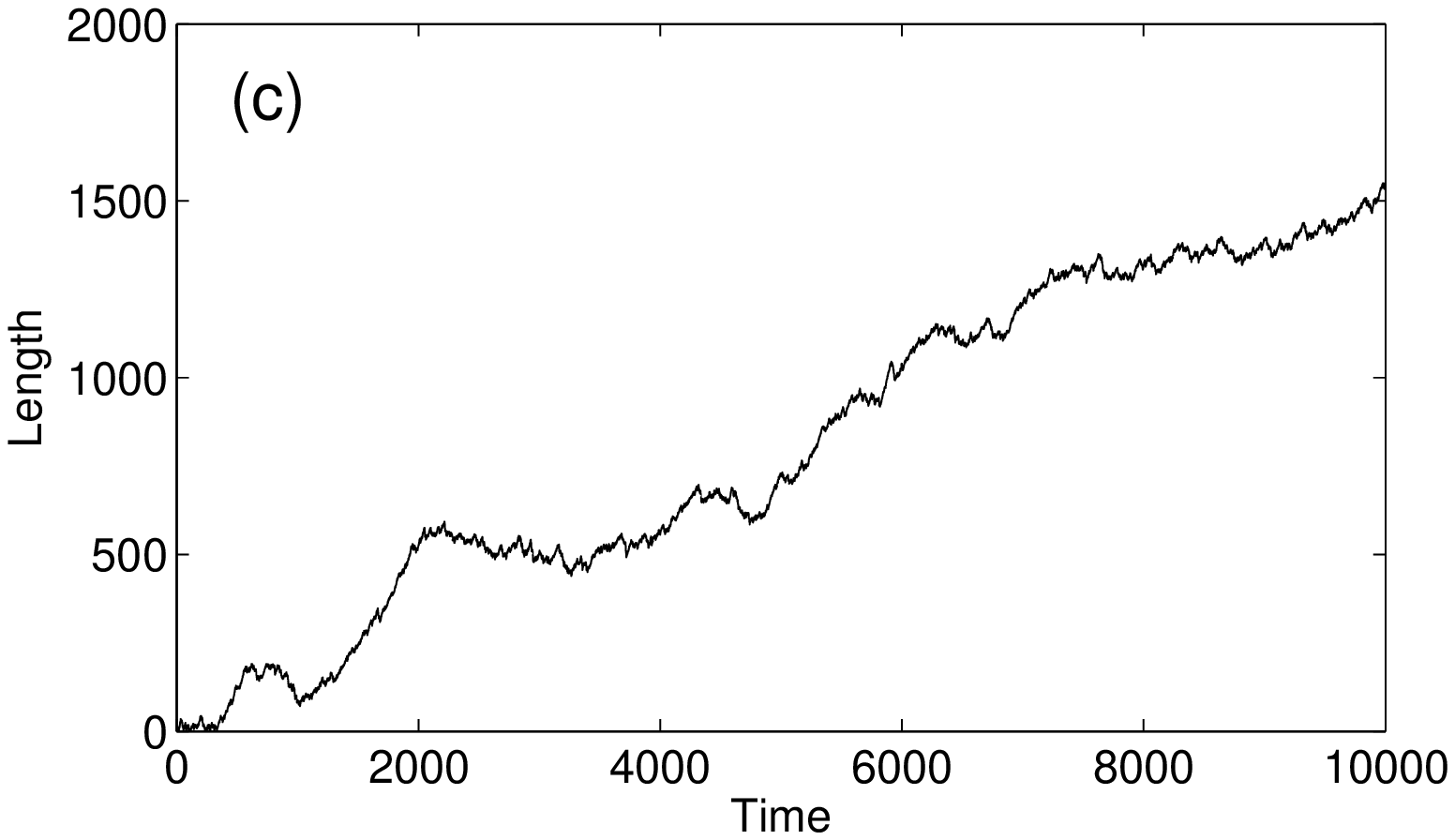}
\caption{Numerical simulations of typical microtubule lengths versus time for
  detachment rate $\mu=5$ and attachment rates: (a) $\lambda=1.4,$ where the
  microtubule generally remains short, (b) $\lambda=1.5$, where the length
  fluctuates strongly, and (c) $\lambda= 1.6$, where the microtubule grows
  nearly steadily.}
\label{l-vs-t}
\end{figure}

Microtubules are polar linear polymers that perform major organizational
tasks in living cells \cite{FBL,VCJ}.  Through a unique feature of
microtubule assembly, termed dynamic instability \cite{MK}, they function as
molecular machines \cite{AH} that move cellular structures during processes
such as cell reproduction \cite{VCJ,WMD}.  A surprising feature of
microtubules is that they remain out of equilibrium under fixed external
conditions and can undergo alternating periods of rapid growth and even more
rapid shrinking (Fig.~\ref{l-vs-t}).

These sudden polymerization changes are driven by the interplay between
several fundamental processes.  Microtubules grow by the attachment of
guanosine triphosphate tubulin complexes (GTP) at one end \cite{MK,DM}.
Structural studies indicate that the end of a microtubule must consist of a
``cap'' of consecutive GTP monomers \cite{dimer} for growth to continue
\cite{DM}.  Once polymerized, the GTP of this complex can irreversibly
hydrolyze into guanosine diphosphate (GDP).  If all the monomers in the cap
convert to GDP, the microtubule is destabilized and rapid shrinkage ensues by
the detachment of GDP tubulin units.  The competition between GTP attachment
and hydrolysis from GTP to GDP is believed to lead to the dynamic instability
in which the GTP cap hydrolyzes to GDP and then the microtubule rapidly
depolymerizes.  The stochastic attachment of GTP can, however, lead to a
rescue to the growing phase before the microtubule length shrinks to zero
\cite{FBL,DL}.

The origin of this dynamic instability has been actively investigated.  One
avenue of theoretical work on this dynamical instability is based on models
of mechanical stability \cite{JCF,vanburen05,mol}.  For example, a detailed
stochastic model of a microtubule that includes all the thirteen constituent
protofilaments has been investigated in Ref.~\cite{vanburen05}.  By using
model parameters that were inferred from equilibrium statistical physics,
VanBuren et al.\ \cite{vanburen05} found some characteristics of microtubule
evolution that agreed with experimental data \cite{mandelkow91}.  The
disadvantage of this detailed modeling, however, is its complexity, so that
it is generally not possible to develop an intuitive understanding of
microtubule evolution.

Another approach for modeling the dynamics of microtubules is based on
effective two-state models that describe the dynamics in terms of a switching
between a growing and a shrinking state \cite{DL,Bi,HZ,MKMC,MGGA,hill}.  The
essence of many of these models is that a microtubule exists either in a
growing phase (where a GTP cap exists at the end of the microtubule) or a
shrinking phase (without a GTP cap), and that there are stochastic
transitions between these two states.  By tuning parameters appropriately, it
is possible to reproduce the phase changes between the growing and shrinking
phases of microtubules that have been observed experimentally \cite{MK}.
While the two-state model has the advantage of having only a few parameters,
a constant rate of switching between a growing and shrinking microtubule is
built into the model.  Thus switching models cannot account for the
stochastic avalanches and catastrophes that occur in real microtubules.

On the other hand, a minimalist model of microtubule dynamics has been
proposed and investigated by Flyvbjerg et al.\ \cite{F}.  In their model,
they dispense with attempts to capture all of the myriad of experimental
parameters within a detailed model, but instead constructed an effective
continuous theory to describe microtubule dynamics.  Their goal was to
construct an effective theory that contained as few details as possible.  As
stated in Ref.~\cite{F}, they envision that their effective theory should be
derivable from a fundamental, microscopic theory and its parameters.

This minimalist modeling is the approach that we adopt in the present work.
We investigate a recently introduced \cite{ZLSW,CR} kinetic model that
accounts for many aspects of microtubule evolution.  Our main result is that
only a few essential parameters with simple physical interpretations are
needed to describe the rich features of microtubule growth, catastrophes, and
rescues \cite{short}.

We treat a microtubule as a linear polymer that consists of GTP or GDP
monomers that we denote as $+$ and $-$, respectively.  To emphasize this
connection between chemistry and the model, we will write the former as
GTP$^+$ and the latter as GDP$^-$.  The state of a microtubule evolves due to
the following three processes:
\begin{enumerate}
\item Attachment: A microtubule grows by attachment of a guanosine
  triphosphate (GTP$^+$) monomer.
\begin{eqnarray*}
    &|\cdots +\rangle \Longrightarrow |\cdots ++\rangle\qquad {\rm rate}~ \lambda\\
    &~~\,|\cdots -\rangle \Longrightarrow |\cdots -+\rangle\qquad {\rm  rate}~ p\lambda.
\end{eqnarray*}

\item Conversion: Once part of the microtubule, each GTP$^+$ can
  independently convert by hydrolysis to a guanosine diphosphate (GDP$^-$).
\begin{eqnarray*}
 ~~~~~~~~~~~|\cdots +\cdots\rangle \Longrightarrow |\cdots-\cdots \rangle \qquad {\rm rate}~ 1.
\end{eqnarray*}

\item Detachment: a microtubule shrinks due to detachment of a GDP$^-$
  monomer {\em only from the end of the microtubule}.
\begin{eqnarray*}
|\cdots-\,\rangle\Longrightarrow |\cdots\rangle\qquad {\rm rate}~ \mu.
\end{eqnarray*}
\end{enumerate}
Here the symbols $|$ and $\rangle$ denote the terminal and the active end of
the microtubule.  It is worth mentioning that these steps are similar to
those in a recently-introduced model of DNA sequence evolution \cite{MLA},
and that some of the results about the structure of DNA sequences seem to be
related to our results about island size distributions in microtubules.

Generically, the $(\lambda,\mu,p)$ phase space separates into a region where
the microtubule grows (on average) with a certain rate $V(\lambda,\mu,p)$,
and a compact phase where the average microtubule length is finite.  These
two phases are separated by a phase boundary $\mu=\mu_*(\lambda,p)$ along
which the growth rate $V(\lambda,\mu,p)$ vanishes.  While the behavior of a
microtubule for general parameter values is of interest, we will primarily
focus on extreme values of the governing parameters where we can obtain a
detailed statistical characterization of the microtubule structure.  For
certain properties, such as the shape of the phase diagram, we will also
present results of numerical simulations of the model.

In Sec.~\ref{growth}, we study the evolution of a microtubule under
unrestricted growth conditions---namely no detachment and an attachment rate
that does not depend on the identity of the last monomer.  Our results here
are relevant to understanding the distribution of cap length and the
diffusion coefficient of the tip of the microtubule in the growth phase.  The
predictions of the model in this limit could also be useful in understanding
the binding pattern of proteins to microtubules \cite{TGSM}.  Since proteins
are important regulatory factors in microtubule polymerization, these results
could prove useful in interpreting the effects of proteins on microtubule
growth.

\begin{figure}[ht]
\includegraphics*[width=0.4\textwidth]{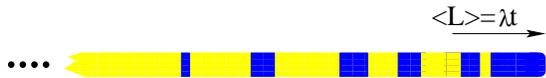}
\caption{(Color online)~ Cartoon of a microtubule in unrestricted growth.
  Regions of GTP$^+$ are shown dark (blue) and regions of GDP$^-$ are light
  (yellow).  The GTP$^+$ regions get shorter further from the tip that
  advances as $\lambda t$, while the GDP$^-$ regions get longer.}
\label{cartoon}
\end{figure}

By a master equation approach, we will determine both the number of GTP$^+$
monomers on a microtubule, as well as the length distributions of GTP$^+$
and GDP$^-$ islands (Fig.~\ref{cartoon}).  Many of these analytical
predictions are verified by numerical simulations.  In
Sec.~\ref{constrained}, we extend our approach to the case of constrained
growth, $p\ne 1$, in which microtubule growth depends on whether the last
monomer is a GTP$^+$ or a GDP$^-$.  In Sec.~\ref{instant}, we investigate the
phenomenon of ``catastrophe'' for infinite detachment rate $\mu$, in which a
microtubule shrinks to zero length when all of its constituent monomers
convert to GDP$^-$.  We derive the asymptotic behavior of the catastrophe
probability by expressing it as an infinite product and recognizing the
connection of this product with modular functions.  We also determine the
asymptotic behavior of the size distribution of avalanches, namely, sequences
of consecutive GDP$^-$ detachment events.  Finally, in Sec.~\ref{general}, we
discuss the behavior of a microtubule for general parameter values through a
combination of numerical and analytic results.  Here numerical simulations
are useful to extract quantitative results for parameter values that are note
amenable to theoretical analysis.  Several calculational details are given in
the appendixes.

\section{UNRESTRICTED GROWTH}
\label{growth}

We define unrestricted growth as the limit of detachment rate $\mu=0$, so
that a microtubule grows without bound.  Here we consider the special case
where the attachment rate does not depend on the identity of the last
monomer; that is, the limit of $p=1$, where the attachment is unconstrained.
Because of the latter condition, the number $N$ of GTP$^+$ monomers decouples
from the number of GDP$^-$, a greatly simplifying feature.

\subsection{Distribution of Positive Monomers}

The average number of GTP$^+$ monomers evolves as
\begin{equation}
\label{N-eq}
\frac{d}{d t}\,\langle N\rangle =\lambda -\langle N\rangle .
\end{equation}
The gain term accounts for the adsorption of a GTP$^+$ at rate $\lambda$,
while the loss term accounts for the conversion events GTP$^+$ $\to$ GDP$^-$,
each of which occurs with rate 1.  Thus $\langle N\rangle$ approaches its
stationary value of $\lambda$ exponentially quickly,
\begin{equation}
\label{N-av}
\langle N\rangle = \lambda (1-e^{-t}).
\end{equation}

More generally, consider the probability $\Pi_N(t)$ that there are $N$ GTP$^+$
monomers at time $t$.  This probability evolves according to
\begin{equation}
\label{PNt}
\frac{d\Pi_N}{dt} = -(N+\lambda)\Pi_N+\lambda\Pi_{N-1}+(N+1)\Pi_{N+1}.
\end{equation}
The loss term $(N+\lambda)\Pi_N$ accounts for conversion events GTP$^+$ $\to$
GDP$^-$ that occur with total rate $N$, and the attachment of a GTP$^+$ at
the end of the microtubule of length $N$ with rate $\lambda$.  The gain terms
can be explained similarly.

In terms of generating function $\Pi(z)\equiv\sum_{N=0}^\infty \Pi_N z^N$,
Eq.~\eqref{PNt} can be recast as the differential equation
\begin{equation}
\label{PNz}
\frac{\partial\Pi}{\partial t} = (1-z)\left(\frac{\partial \Pi}{\partial
      z}-\lambda \Pi\right).
\end{equation}
Introducing $\mathcal{Q}=\Pi e^{-\lambda z}$ and 
$y=\log (1-z)$, we
transform Eq.~\eqref{PNz} into the wave equation
\begin{equation}
\label{wave}
\frac{\partial \mathcal{Q}}{\partial t}+ \frac{\partial \mathcal{Q}}{\partial y}=0,
\end{equation}
whose solution is an arbitrary function of $t-y$ or, equivalently,
$(1-z)e^{-t}$.  If the system initially is a microtubule of zero length,
$\Pi_N(t=0)=\delta_{N,0}$, the initial generating function $\Pi(z,t=0)=1$, so
that $\mathcal{Q}=e^{-\lambda z}= e^{\lambda(1-z)} e^{-\lambda}$.  Thus for
$t>0$, $\mathcal{Q}= e^{\lambda(1-z)e^{-t}} e^{-\lambda}$, from which
\begin{equation}
\label{Poisson}
\Pi(z,t) = e^{-\lambda(1-z)(1-e^{-t})}.
\end{equation}
Expanding this expression in a power series in $z$, the probability for the
system to contain $N$ GTP$^+$ monomers is the time-dependent Poisson distribution
\begin{equation}
\label{Poisson-t}
\Pi_N(t) =  \frac{[\lambda(1-e^{-t})]^N}{N!}\,e^{-\lambda(1-e^{-t})}.
\end{equation}
From this result, the mean number of GTP$^+$ monomers and its variance are
\begin{equation}
\label{N-av-var}
\langle N\rangle = \langle N^2\rangle - \langle N\rangle^2 = \lambda(1-e^{-t}).
\end{equation}

\subsection{Tubule Length Distributions}

The length distribution $P(L,t)$ of the microtubule evolves according to the
master equation
\begin{equation}
\label{F}
\frac{d P(L,t)}{d t} = \lambda\left[P(L-1,t)-P(L,t)\right]
\end{equation}
For the initial condition $P(L,0)=\delta_{L,0}$, the solution is again the
Poisson distribution
\begin{equation}
\label{F-sol}
P(L,t) = \frac{(\lambda t)^L}{L!}\,e^{-\lambda t}
\end{equation}
{}from which the average and the variance are
\begin{equation}
\label{L-av-var}
\langle L\rangle=\lambda t,\quad \langle L^2\rangle-\langle L\rangle^2=\lambda t\,.
\end{equation}
Thus the growth rate of the microtubule and the diffusion coefficient of the
tip are
\begin{equation}
\label{VD-simple}
V=\lambda, \quad D=\lambda/2
\end{equation}

A more comprehensive description is provided by the joint distribution
$P(L,N,t)$ that a microtubule has length $L$ and contains $N$ GTP$^+$
monomers at time $t$.  This distribution evolves as
\begin{eqnarray}
\label{PLN}
\frac{d P(L,N)}{d t} &=& \lambda P(L-1,N-1)-(N+\lambda)P(L,N) \nonumber \\
&+&(N+1)P(L,N+1).
\end{eqnarray}
This joint distribution does {\em not} factorize, that is, $P(L,N,t)\ne
P(L,t)\,\Pi_N(t)$, because $\langle LN\rangle\ne \langle L\rangle \langle
N\rangle$.  To demonstrate this inequality, we compute $\langle LN\rangle$ by
multiplying Eq.~\eqref{PLN} by $LN$ and summing over all $L\geq N\geq 0$ to
give
\begin{equation}
\label{LN}
\frac{d}{dt}\,\langle LN\rangle=\lambda(\langle L\rangle+\langle N\rangle+1)
-\langle LN\rangle.
\end{equation}
Using Eqs.~\eqref{N-av-var} and \eqref{L-av-var} for $\langle N\rangle$ and
$\langle L\rangle$ and integrating we obtain
\begin{eqnarray}
\label{LN-av}
\langle LN\rangle &=& \lambda^2t\,(1-e^{-t})+\lambda(1-e^{-t})\nonumber\\
&=&\langle L\rangle \langle N\rangle +\langle N \rangle.
\end{eqnarray}
Using Eq.~\eqref{L-av-var}, we have $\langle
LN\rangle= \langle L\rangle \langle N\rangle(1+ \frac{1}{\lambda t})$, so
that the joint distribution is factorizable asymptotically.  For
completeness, we give the full solution for $P(L,N,t)$ in Appendix
\ref{app-P}.

\subsection{Cap Length Distribution}
\label{sec:cld}

\begin{figure}
\includegraphics*[width=0.4\textwidth]{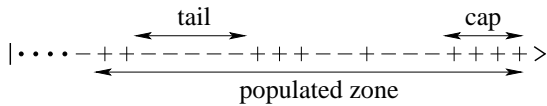}
\caption{Representative configuration of a microtubule, with a GTP$^+$ cap of
  length 4, then three GTP$^+$ islands of lengths 1, 3, and 2, and three
  GDP$^-$ islands of lengths 3, 2, and a ``tail'' of length 5.  The rest of
  the microtubule consists of GDP$^-$.}
\label{cap-fig}
\end{figure}

Because of the conversion process GTP$^+$ $\to$ GDP$^-$, the tip of the
microtubule is comprised predominantly of GTP$^+$, while the tail exclusively
consists of GDP$^-$.  The region from the tip until the first GDP$^-$ monomer
is known as the {\em cap} (Fig.~\ref{cap-fig}) and it plays a fundamental
role in microtubule function.  We now use the master equation approach to
determine the cap length distribution.

\begin{figure}[ht]
\includegraphics*[width=0.8\columnwidth]{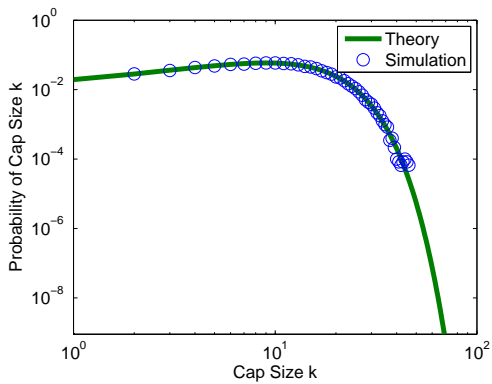}
\caption{Cap length distribution obtained from simulations at $\mu=0$,
  $\lambda=\ 100$, and $p=1$ compared to the theoretical prediction of
  Eq.~\eqref{nk-sol}.}
\label{capsize}
\end{figure}

Consider a cap of length $k$.  Its length increases by 1 due to the
attachment of a GTP$^+$ at rate $\lambda$.  The conversion of any GTP$^+$
into a GDP$^-$ at rate 1 reduces the cap length from $k$ to an arbitrary
value $s<k$.  These processes lead to the following master equation for the
probability $n_k$ that the cap length equals $k$:
\begin{equation}
\label{nk}
\dot n_k = \lambda(n_{k-1}-n_k)-kn_k+\sum_{s\geq k+1}n_s.
\end{equation}
Equation~\eqref{nk} is also valid for $k=0$ if we set $n_{-1}\equiv 0$. Note
that $n_0\equiv {\rm Prob}\{-\rangle\}$ is the probability for a cap of
length zero. We now solve for the stationary distribution by summing the
first $k-1$ of Eqs.~\eqref{nk} with $\dot n_k$ set to zero to obtain
\begin{equation}
 n_{k-1} = \frac{k}{\lambda} \sum_{s\geq k}n_s.
\end{equation}
The cumulative distribution, $ N_k=\sum_{s\geq k}n_s$, thus satisfies the
recursion
\begin{equation}
\label{Nk}
 N_k = \frac{\lambda}{k+\lambda} N_{k-1}.
\end{equation}
Using the normalization $N_0=1$ and iterating, we obtain the solution in
terms of the Gamma function \cite{AS}:
\begin{equation}
\label{Nk-soln}
 N_k = \frac{\lambda^k\, \Gamma(1+\lambda)}{\Gamma(k+1+\lambda)}.
\end{equation}
Hence the cap length distribution is
\begin{equation}
\label{nk-sol}
 n_k = \frac{\Gamma(1\!+\!\lambda)}{\Gamma(k\!+\!2\!+\!\lambda)}\, (k+1)\lambda^k
\end{equation}
and the first few terms are
\begin{subequations}
\begin{align}
 n_0 &= \frac{1}{1+\lambda}\nonumber \\
 n_1 &= \frac{2\lambda}{(1+\lambda)(2+\lambda)\nonumber}\\
 n_2 &= \frac{3\lambda^2}{(1+\lambda)(2+\lambda)(3+\lambda)}.\nonumber
\end{align}
\end{subequations}
Results of direct simulation of the kinetic model are compared to the
predicted cap length distribution (Eq.~\eqref{nk-sol}) in Fig.~\ref{capsize}.
Because of the finite length of the simulated microtubule, there is a largest
cap length that is accessible numerically.  Aside from this limitation, the
simulations results are in agreement with theoretical predictions.

It is instructive to determine the dependence of the average cap length
$\langle k\rangle =\sum_{k\geq 0}kn_k$ on $\lambda$.  Using
$n_k=N_k-N_{k+1}$, we rearrange $\langle k\rangle$ into
\begin{equation}
\label{cap}
\langle k\rangle =\sum_{k\geq 1}N_k
\end{equation}
Using \eqref{Nk-soln}, the above sum may be written in terms of the confluent
hypergeometric series \cite{AS}:
\begin{equation}
\label{cap-av}
\langle k\rangle =-1+F(1;1+\lambda; \lambda).
\end{equation}
We now determine the asymptotic behavior of $\langle k\rangle$ by using the
integral representation
\begin{equation*}
 F(a;b; z)=\frac{\Gamma(b)}{\Gamma(b-a)\,\Gamma(a)}
\int_0^1 dt\,e^{zt}\,t^{a-1}(1-t)^{b-a-1}
\end{equation*}
to recast the average cap length \eqref{cap-av} as
\begin{equation}
\label{cap-average}
\langle k\rangle =-1+ \lambda \left(\frac{e}{\lambda}\right)^{\lambda} 
\gamma(\lambda,\lambda),
\end{equation}
where $\gamma(a,x)=\int_0^x dt\,t^{a-1} e^{-t}$ is the (lower) incomplete
gamma function.

In the realistic limit of $\lambda\gg 1$, we use the large $\lambda$
asymptotics
\begin{equation*}
\gamma(\lambda,\lambda)\to \frac{1}{2}\,\Gamma(\lambda), \quad 
\Gamma(\lambda)\sim \sqrt{\frac{2\pi}{\lambda}}\left(\frac{\lambda}{e}\right)^{\lambda},
\end{equation*}
to give
\begin{equation}
\label{kav-large}
\langle k\rangle\to \sqrt{\pi\lambda/2}\quad {\rm as}\quad \lambda\to\infty
\end{equation}
Thus even though the number of GTP$^+$ monomers equals $\lambda$, only
$\sqrt{\lambda}$ of them comprise the microtubule cap, as qualitatively
illustrated in Fig.~\ref{cartoon}.  Note that the average cap length is
proportional to the square-root of the velocity; essentially the same result
was obtained from the coarse-grained theory of Flyvbjerg et al.\ \cite{F}.

\subsection{Island Size Distributions}

At a finer level of resolution, we determine the distribution of island sizes
at the tip of a microtubule (Fig.~\ref{cap-fig}).  A simple characteristic of this
population is the average number $I$ of GTP$^+$ islands.  If all GTP$^+$ islands were
approximately as long as the cap, we would have $I\sim \langle
N\rangle/\langle k\rangle \sim \sqrt{\lambda}$.  As we shall see, however,
$I$ scales linearly with $\lambda$ because most islands are short.  A similar
dichotomy arises for negative islands.

To write the master equation for the average number of islands, note that the
conversion GTP$^+$ $\to$ GDP$^-$ eliminates islands of size 1.  Additionally, an
island of size $k\geq 3$ splits into two daughter islands, and hence the
number of islands increases by one, if conversion occurs at any one of the
$k-2$ in the interior of an island as illustrated below:
\begin{equation*}
-+\underbrace{+\cdots +}_{k-2} +-\,.
\end{equation*}
Conversely, if the cap has length 0, attachment creates a new cap of length 1
at rate $\lambda$.  The net result of these processes is encoded in the rate
equation
\begin{equation}
\label{island-eq}
\frac{d I}{dt} = \sum_{k\geq 1}(k-2)I_k +\lambda n_0,
\end{equation}
with $I_k$ the average number of GTP$^+$ islands of size $k$.

We now use the sum rules $I = \sum_{k\geq 1}I_k$ and $\langle N\rangle =
\sum_{k\geq 1}k I_k$ to recast \eqref{island-eq} as
\begin{equation}
\label{island-main}
\frac{d I}{dt} = \langle N\rangle - 2I +\lambda n_0
\end{equation}
from which the steady-state average number of islands is
\begin{equation}
\label{island-sol}
I = \frac{\langle N\rangle +\lambda n_0}{2}
= \frac{\lambda}{2}\,\frac{2+\lambda}{1+\lambda}.
\end{equation}
For large $\lambda$, the number of islands approaches $\lambda/2$, while the
number of GTP$^+$ monomers equals $\lambda$.  Thus the typical island size is 2.
Nevertheless, as we now show, the GTP$^+$ and GDP$^-$ island distributions actually have
power-law tails, with different exponents for each species.

The GTP$^+$ island size distribution evolves according to the master equation
\begin{equation}
\label{Ik-eq}
\dot I_k = -kI_k+2\sum_{s\geq k+1}I_s +\lambda(n_{k-1}-n_k)
\end{equation}
This equation is similar in spirit to Eq.~\eqref{nk} for the cap length
distribution.  As a useful self-consistency check, the sum of
Eqs.~\eqref{Ik-eq} gives \eqref{island-main}, while multiplying \eqref{Ik-eq}
by $k$ and summing over all $k\geq 1$ gives Eq.~\eqref{N-eq}.

\begin{figure}[ht]
\includegraphics*[width=0.9\columnwidth]{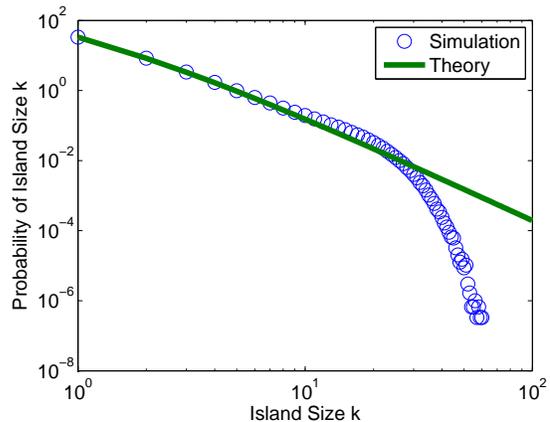}
\caption{Simulation results at $\mu=0$, $\lambda=100$, and $p=1$ for the size
  distribution of positive islands, $I_{k}/\lambda$.  The solid line is the
  theoretical prediction of Eq.~\eqref{Ik-sol}.}
\label{islandsize}                                                                 
\end{figure}

The stationary distribution satisfies
\begin{equation}
\label{Ik}
kI_k=2\sum_{s\geq k+1}I_s +\lambda(n_{k-1}-n_k).
\end{equation}
Using $\sum_{s\geq 2}I_s=I-I_1$, we transform \eqref{Ik} at $k=1$ to
\begin{equation*}
\label{I1-eq}
3I_1=2I+\lambda(n_0-n_1)
\end{equation*}
Similarly, using $\sum_{s\geq 3}I_s=I-I_1-I_2$ we transform \eqref{Ik} at
$k=2$ to
\begin{equation*}
\label{I2-eq}
4I_2=2(I-I_1)+\lambda(n_1-n_2)
\end{equation*}
Thus using \eqref{nk-sol} and
\eqref{Ik} we obtain
\begin{subequations}
\begin{align}
\label{I1}
 I_1 &= \frac{\lambda}{3}+\frac{4\lambda}{3(1+\lambda)(2+\lambda)}\\
\label{I2}
 I_2 &= \frac{\lambda}{12}
 +\frac{25\lambda^2-6\lambda}{12(1+\lambda)(2+\lambda)(3+\lambda)}.
\end{align}
\end{subequations}
The same procedure gives $I_k$ for larger $k$.  

Since the $I_k$ represent the {\em average} number of islands of size $k$,
they become meaningful only for $\lambda\to\infty$ where an appreciable
number of such islands exist.  In this limit, we write $I_k$ more compactly
by first rearranging \eqref{Ik} into the equivalent form
\begin{equation}
\label{Ikk}
(k-1)I_{k-1}-(k+2)I_k=\lambda(n_{k-2}-2n_{k-1}+n_k).
\end{equation}
We  then use \eqref{nk-sol} and the asymptotic properties of the Gamma
function to find that the right-hand side of  Eq.~\eqref{Ikk} is
\begin{equation*}
\lambda(n_{k-2}-2n_{k-1}+n_k)=-\frac{3k+1}{\lambda}+O\left(\frac{1}{\lambda^2}\right),
\end{equation*}
and is therefore negligible in the large-$\lambda$ limit.  Thus \eqref{Ikk}
reduces to $(k-1)I_{k-1}=(k+2)I_k$, with solution $I_k=A/[k(k+1)(k+2)]$.  We
find the amplitude $A$ by matching with the exact result, Eq.~\eqref{I1}, to
give $I_1=\lambda/3$ for large $\lambda$.  The final result is
\begin{equation}
\label{Ik-sol}
I_k=\frac{2\lambda}{k(k+1)(k+2)}.
\end{equation}
In the large $\lambda$ limit, $I=\lambda/2$, and the above result can be re-written as
\begin{equation}
\label{Ik-I}
\frac{I_k}{I}=\frac{4}{k(k+1)(k+2)}.
\end{equation}
Remarkably, the size distribution of the positive islands is {\em
  identical\/} to the degree distribution of a growing network with strictly
linear preferential attachment \cite{network1,network2,network3}.

The results for the island size distribution in the large $\lambda$ limit are
compared to simulation results in Fig.~\ref{islandsize}.  These asymptotic
results are expected to apply to island sizes $k$ much smaller than the size
of the cap which scales as $\sqrt \lambda$.  The distributions obtained from
the numerical simulations should then obey the theoretical form but with a
finite-size cutoff.  The results in Fig.~\ref{islandsize} are consistent with
this picture but, interestingly, the numerical distribution rises above the
theoretical curve before falling sharply below it.  This anomaly occurs in
many heterogeneous growing network models, and it can be fully characterized
in terms of finite-size effects \cite{finite}.

\subsection{Continuum Limit, $\lambda\to\infty$}
\label{continuum}

When $\lambda\to\infty$, both the length of the cap and the length of the
region that contains GTP$^+$ become large.  In this limit, the results from
the discrete master equation can be expressed much more elegantly and
completely by a continuum approach.  The fundamental feature is that the
conversion process GTP$^+$ $\to$ GDP$^-$ occurs independently for each
monomer.  Since the residence time of each monomer increases linearly with
distance from the tip, the probability that a GTP$^+$ does not convert decays
exponentially with distance from the tip.  This fact alone is sufficient to
derive all the island distributions.

Consider first the length $\ell$ of the populated region
(Fig.~\ref{cap-fig}).  For a GTP$^+$ that is a distance $x$ from the tip, its
residence time is $\tau=x/\lambda$ in the limit of large $\lambda$.  Thus the
probability that this GTP$^+$ does not convert is $e^{-\tau}=e^{-x/\lambda}$.
We thus estimate $\ell$ from the extremal criterion \cite{extremal}
\begin{equation}
\label{ell}
1=\sum_{x\geq \ell} e^{-x/\lambda}=(1-e^{-1/\lambda})^{-1}e^{-\ell/\lambda},
\end{equation}
that merely states that there is of the order of a single GTP$^+$ further
than a distance $\ell$ from the tip.  Since $(1-e^{-1/\lambda})^{-1}\to
\lambda$ when $\lambda$ is large, the length of the active region scales as
\begin{equation}
\label{ell-sol}
\ell=\lambda\ln\lambda
\end{equation}

The probability that the cap has length $k$ is given by
\begin{equation*}
(1-e^{-(k+1)/\lambda})\prod_{j=1}^k e^{-j/\lambda}.
\end{equation*}
The product ensures that all monomers between the tip and a distance $k$ from
the tip are GTP$^+$, while the prefactor gives the probability that a monomer
is a distance $k+1$ from the tip is a GDP$^-$.  Expanding the prefactor for
large $\lambda$ and rewriting the product as the sum in the exponent, we
obtain
\begin{equation}
\label{nk-cont}
n_k\sim \frac{k+1}{\lambda}\,\, e^{-k(k+1)/2\lambda},
\end{equation}
a result that also can be obtained by taking the large-$\lambda$ limit of
the exact result for $n_k$ given in Eq.~\eqref{nk-sol}.

Similarly, the probability to find a positive island of length $k$ that
occupies sites $x+1,x+2,\ldots,x+k$ is
\begin{equation}
\label{isl-x}
(1-e^{-x/\lambda})(1-e^{-(x+k+1)/\lambda})\prod_{j=1}^k e^{-(x+j)/\lambda}.
\end{equation}
The two prefactors ensure that sites $x$ and $x+k+1$ consist of GDP$^-$,
while the product ensures that all sites between $x+1$ and $x+k$ are GTP$^+$.

Most islands are far from the tip and they are relatively short, $k\ll x$, so
that \eqref{isl-x} simplifies to
\begin{equation}
\label{isl-x-simple}
(1-e^{-x/\lambda})^2 e^{-kx/\lambda}\, e^{-k^2/2\lambda}.
\end{equation}
The total number of islands of length $k$ is obtained by summing the island
density \eqref{isl-x-simple} over all $x$.  Since $\lambda\gg 1$, we replace
the summation by integration and obtain
\begin{eqnarray}
\label{isl-k}
I_k&=&\int_0^\infty \!\!dx\,(1-e^{-x/\lambda})^2\, e^{-kx/\lambda}\,
e^{-k^2/2\lambda}\nonumber \\
&=&\frac{2\lambda}{k(k\!+\!1)(k\!+\!2)}\, e^{-k^2/2\lambda}.
\end{eqnarray}
The power law tail agrees with Eq.~\eqref{Ik-sol}, whose derivation
explicitly invoked the $\lambda\to\infty$ limit.

We can also obtain the density of negative islands in this continuum
description, a result that seems impossible to derive by a microscopic master
equation description.  In parallel with \eqref{isl-x-simple}, the density of
negative islands of length $k\ll x$ with one end at $x$ is given by
\begin{equation}
\label{isl-negative}
e^{-2x/\lambda} (1-e^{-x/\lambda})^k, 
\end{equation}
and the total number of negative islands of length $k$ is
\begin{equation}
\label{isl-k-negative}
J_k=\int_0^\infty dx\,e^{-2x/\lambda} (1-e^{-x/\lambda})^k =\frac{\lambda}{(k+1)(k+2)}.
\end{equation}
Again, we find a power-law tail for the GDP$^-$ island size distribution, but
with exponent 2.  The total number of GDP$^-$ monomers within the populated
zone is then $\sum_{k\geq 1}kJ_k$.  While this sum formally diverges, we use
the upper size cutoff, $k_*\sim \lambda$ to obtain $\sum_{k\geq 1}kJ_k\simeq
\lambda\ln\lambda$.  Since the length of the populated zone $\ell\sim
\lambda\ln \lambda$,this zone therefore predominantly consists of GDP$^-$
islands.

In analogy with the cap, consider now the ``tail''---the last island of
GDP$^-$ within the populated zone (see Fig.~\ref{cap-fig}).  The probability
$m_k$ that it has length $k$ is
\begin{equation}
\label{last}
m_k = e^{-\ell/\lambda} (1-e^{-\ell/\lambda})^k.
\end{equation}
Using \eqref{ell-sol} we simplify the above expression to
\begin{equation*}
\label{last-simple}
m_k = \lambda^{-1} (1-\lambda^{-1})^k = \lambda^{-1} e^{-k/\lambda}.
\end{equation*}
Hence the average length of the tail is
\begin{equation}
\label{last-length}
\langle k\rangle = \sum_{k\geq 1} km_k = \lambda,
\end{equation}
which is much longer (on average) than the cap.

\section{CONSTRAINED GROWTH}
\label{constrained}

When $p\ne 1$, the rate of attachment depends on the state of the tip of the
microtubule---attachment to a GTP$^+$ occurs with rate $\lambda$ while
attachment to a GDP$^-$ occurs with rate $p\lambda$.  While this state
dependence makes the master equation description for the properties of the
tubule more complicated, qualitative features about the structure of the
populated zone are the same as those in the case $p=1$.  In this section, we
outline some of the basic features of the populated zone when $p\ne 1$, but
we still keep $\mu=0$.

\subsection{Distribution of GTP$^+$}

The average number of GTP$^+$ monomers now evolves according to the rate equation
\begin{equation}
\label{N-avt}
 \frac{d}{dt}\, \langle N\rangle = -\langle N\rangle+p\lambda n_0+\lambda (1-n_0),
\end{equation}
which should be compared to the rate equation Eq.~\eqref{N-eq} for the case
$p=1$.  The loss term on the right-hand side describes the conversion
GTP$^+\to $GDP$^-$, while the remaining terms represent gain due to
attachment to a GTP$^+$ with rate $\lambda$ and to a GDP$^-$ with rate
$p\lambda$.  Here $n_0$ is the probability for a cap of length zero, that is,
the last site is a GDP$^-$.  The stationary solution to \eqref{N-avt} is
\begin{equation}
\label{N-av-inf}
 \langle N\rangle = p\lambda n_0+\lambda (1-n_0),
\end{equation}
so we need to determine $n_0$.  By extending Eq.~\eqref{nk} to the case $p\ne
1$, we then find that $n_0$ is governed by the rate equation
\begin{equation}
\label{n0-rate}
\dot n_0 = -p\lambda n_0+(1-n_0).
\end{equation}
Thus asymptotically $n_0 = \frac{1}{1+p\lambda}$ and substituting into
\eqref{N-av-inf}, the average number of GTP$^+$ monomers is
\begin{equation}
\label{N-av-sol}
\langle N\rangle=p\lambda\,\frac{1+\lambda}{1+p\lambda}
\end{equation}

More generally, we can determine the distribution of the number of GTP$^+$
monomers; the details of this calculation are presented in
Appendix~\ref{app}.

\subsection{Growth Rate and Diffusion Coefficient}
\label{XY-sub}

The growth rate of a microtubule equals $p\lambda$ when the cap length is
zero and to $\lambda$ otherwise.  Therefore
\begin{equation}
\label{Vn}
V(p,\lambda) = p\lambda n_0+\lambda(1-n_0) = p\lambda\,\frac{1+\lambda}{1+p\lambda}
\end{equation}
For the diffusion coefficient of the tip of a microtubule, we need its
mean-square position.  As in the case $p=1$, it is convenient to determine
the probability distribution for the tip position.  Thus we introduce
$X(L,t)$ and $Y(L,t)$, the probabilities that the microtubule length equals
$L$ and the last monomer is a GTP$^+$ or a GDP$^-$, respectively.  These
probabilities satisfy
\begin{subequations}
\begin{align}
\label{X}
   \frac{dX(L)}{d t}&= \lambda X(L\!-\!1)+p\lambda Y(L\!-\!1)-(1\!+\!\lambda)X(L)\\
\label{Y} 
   \frac{d Y(L)}{d t}&= X(L) - p\lambda Y(L),
\end{align} 
\end{subequations}
Summing these equations, the length distribution $P(L)=X(L)+Y(L)$ satisfies
\begin{equation}
\label{Fp}
   \frac{dP(L)}{d t} =\lambda X(L-1)+p\lambda Y(L-1)-\lambda X(L)-p\lambda Y(L).
\end{equation}

The state of the last monomer does not depend on the microtubule length $L$
for large $L$.  Thus asymptotically 
\begin{equation}
\label{XY}
X(L)=(1-n_0)P(L), \quad Y(L)=n_0 P(L).
\end{equation}
Substituting \eqref{XY} into \eqref{Fp} we obtain a master equation for the
tubule length distribution of the same form as Eq.~\eqref{F}, but with
prefactor $V$ given by \eqref{Vn} instead of $\lambda$.  As a result of this
correspondence, we infer that the diffusion coefficient is one-half of the
growth rate,
\begin{equation}
\label{D-sol}
D(p,\lambda)=\frac{1}{2}\,p\lambda\,\frac{1+\lambda}{1+p\lambda}.
\end{equation}
For large $\lambda$ both the growth rate of the tip of the microtubule and
its diffusion coefficient approach the corresponding expressions in
Eq.~\eqref{VD-simple} for the case $p=1$.

\subsection{Cap Length Distribution}

The master equations for the cap length distribution are the same as in the
$p=1$ case when $k\geq 2$.  The master equations for $k=0$ and $k=1$ are
slightly changed to account for the different rates at which attachment
occurs at a GDP$^-$ monomer:
\begin{eqnarray*}
 p\lambda n_0&=&N_{1}=1-n_0\\
 (1+\lambda)n_1-p\lambda n_0&=&N_2=1-n_0-n_1
\end{eqnarray*}
Solving iteratively we recover $n_0= \frac{1}{1+p\lambda}$ and also obtain
\begin{subequations}
\begin{align}
\label{n1-p}
 n_1 &= \frac{2p\lambda}{(1+p\lambda)(2+\lambda)}\\
\label{n2-p}
 n_2 &= \frac{3p\lambda^2}{(1+p\lambda)(2+\lambda)(3+\lambda)},
 \end{align}
\end{subequations} {\it etc.}  The general solution for the $n_k$ is found by
the same method as in Sec.~\ref{sec:cld} to be
\begin{equation}
\label{nkp}
 n_k = (k+1)\lambda^k\,\frac{p}{1+p\lambda}\, 
 \frac{\Gamma(2+\lambda)}{\Gamma(k+2+\lambda)},
\end{equation}
which are valid for $k\geq 1$.  With this length distribution, the average
cap length is then
\begin{equation}
\label{cap-p}
\langle k\rangle = \frac{p}{1+p\lambda}\sum_{k\geq 1} k(k+1)\lambda^k\,
 \frac{\Gamma(2+\lambda)}{\Gamma(k+2+\lambda)},
\end{equation}
and the sum can again be expressed in terms of hypergeometric series as in
Eq.~\eqref{cap-av}.  Rather than following this path, we focus on the most
interesting limit of large $\lambda$.  Then the cap length distribution
\eqref{nkp} approaches to the previous solution \eqref{nk-sol} for the case
$p=1$ and the mean length reduces to \eqref{kav-large}, independent of $p$.

\subsection{Island Size Distribution}

For the distribution of island sizes, the master equation remains the same as
in the $p=1$ case when $k\geq 2$.  However, when $k=1$, the master equation
becomes
\begin{equation}
\label{I1-eq-a}
I_1=2\sum_{s\geq 2}I_s +p\lambda n_0 - \lambda n_1
\end{equation}
in the stationary state. Then the average number of islands and the average
number of islands of size 1 are found from
\begin{eqnarray*}
2I &=& \langle N\rangle +p\lambda n_0\\
3I_1 &=& 2I +p\lambda n_0 - \lambda n_1
\end{eqnarray*}
Using $n_0 = \frac{1}{1+p\lambda}$ and Eqs.~\eqref{N-av-sol} and \eqref{n1-p} we obtain
\begin{subequations}
\begin{align}
\label{I-p}
 I &= \frac{p\lambda}{2}\,\frac{2+\lambda}{1+p\lambda}\\
\label{I1-p}
 I_1 &= \frac{p\lambda}{1+p\lambda}
 \left[\frac{\lambda}{3}+\frac{2+\lambda/3}{2+\lambda}\right]
 \end{align}
\end{subequations}

Again, in the limit of large $\lambda$, the average island size distribution
reduces to our previously-quoted results in \eqref{Ik-sol} or equivalently
\eqref{Ik-I}.  The leading behavior in the $\lambda\to\infty$ limit is again
independent of $p$.

\section{Instantaneous Detachment}
\label{instant}

For $\mu>0$, a microtubule can recede if its tip consists of GDP$^-$.  The
competition between this recession and growth by the attachment of GTP$^+$
leads to a rich dynamics in which the microtubule length can fluctuate wildly
under steady conditions.  In this section, we focus on the limiting case of
infinite detachment rate, $\mu=\infty$.  In this limit, any GDP$^-$
monomer(s) at the tip of a microtubule are immediately removed.  Thus the the
tip is always a GTP$^+$; this means that the parameter $p$ become immaterial.
Finally, for $\mu=\infty$, we also require the growth rate $\lambda\to\infty$
to have a microtubule with an appreciable length.  This is the limit
considered below.

As soon as the last monomer of the tubule changes from a GTP$^+$ to a
GDP$^-$, a string of $k$ contiguous GDP$^-$ monomers exist at the tip and
they detach immediately.  We term such an event an {\em avalanche of size
  $k$}.  We now investigate the statistical properties of these avalanches.

\subsection{Catastrophes}
The switches from a growing to a shrinking state of a microtubule are called
{\em catastrophes} \cite{DL}.  If a newly-attached GTP$^+$ at the tip
converts to a GDP$^-$ and the rest of the microtubule consists only of
GDP$^-$ at that moment, the microtubule instantaneously shrinks to zero
length, a phenomenon that can be termed a {\em global catastrophe}.  We now
determine the probability for such a catastrophe to occur.  Formally, the
probability of a global catastrophe is
\begin{equation}
\label{sweep}
\mathcal{C}(\lambda)=\frac{1}{1+\lambda}\prod_{n=1}^\infty (1-e^{-n/\lambda}).
\end{equation}
The factor $(1+\lambda)^{-1}$ gives the probability that the monomer at the
tip converts to a GDP$^-$ before the next attachment event, while the product
gives the probability that the rest of the microtubule consists of GDP$^-$.
In principle, the upper limit in the product is set by the microtubule
length.  However, for $n>\lambda$, each factor in the product is close to 1
and the error made in extending the product to infinity is small.
The expression within the product is obtained under the assumption that the
tubule grows steadily between these complete catastrophes and the smaller,
local catastrophes, are therefore ignored in this calculation.

The leading asymptotic behavior of the infinite product in \eqref{sweep} is
found by expressing it in terms of the Dedekind $\eta$ function
\cite{apostol}
\begin{equation}
\label{eta}
\eta(z)=e^{i\pi z/12}\prod_{n=1}^\infty (1-e^{2\pi i nz}),
\end{equation}
and using a remarkable identity satisfied by this function,
\begin{equation*}
\label{ident}
\eta(-1/z)=\sqrt{-i z}\,\eta(z).
\end{equation*}
For our purposes, we define $a=-i\pi z$ to rewrite this identity as
\begin{equation*}
\label{identity}
\prod_{n=1}^\infty (1-e^{-2an})=\sqrt{\frac{\pi}{a}}\,e^{(a-b)/12}
\prod_{n=1}^\infty (1-e^{-2bn})
\end{equation*}
where $b=\pi^2/a$.  Specializing to the case $a=(2\lambda)^{-1}$ yields
\begin{eqnarray}
\label{S}
\mathcal{C}(\lambda)&=&\frac{\sqrt{2\pi\lambda}}{1+\lambda} \,\,e^{-\pi^2\lambda/6}\, e^{1/24\lambda}\,
\prod_{n\geq 1} (1-e^{-4\pi^2\lambda  n})\nonumber \\
&\sim& \sqrt{\frac{2\pi}{\lambda}}\,\,e^{-\pi^2\lambda/6}.
\end{eqnarray}
Since the time between catastrophes scales as the inverse of the occurrence
probability, this inter-event time becomes very long for large $\lambda$.

\subsection{Avalanche Size Distribution}

In the instantaneous detachment limit, $\mu=\infty$, the catastrophes are
avalanches whose size is determined by the number of GDP$^-$s between the tip
and the first GTP$^+$ island.  A global catastrophe is an avalanche of size
equal to the length of the tubule, whose occurrence probability was
calculated in the preceding section.  Similar arguments can be used to
calculate the size distribution of the smaller avalanches.

Since the cap is large when $\lambda$ is large, an avalanche of size 1 arises
only through the reaction scheme
\begin{equation*}
  |\cdots ++\rangle \Longrightarrow |\cdots +-\rangle
  \Longrightarrow |\cdots +\rangle,
\end{equation*}
where the first step occurs at rate 1 and the second step is instantaneous.
Since attachment proceeds with rate $\lambda$, the probability that
conversion occurs before attachment is $A_1=(1+\lambda)^{-1}$; this expression
gives the relative frequency of avalanches of sizes $\geq 1$.
Analogously, an avalanche of size two is formed by the events
\begin{equation*}
|\cdots +++\rangle \Longrightarrow |\cdots+-+\rangle
 \Longrightarrow |\cdots+--\rangle \Longrightarrow |\cdots+\rangle,
\end{equation*}
and the probability that the first two steps occur before an attachment event
is $A_2\sim \lambda^{-2}$ to lowest order.  At this level of approximation, the
relative frequency of avalanches of size equal to 1 is $A_1-A_2\sim
\lambda^{-1}$.  Since we are interested in the regime $\lambda\gg 1$, we
shall only consider the leading term in the avalanche size distribution.

Generally an avalanche of size $k$ is formed if the system starts in the
configuration
\begin{equation*}
|+\underbrace{+\cdots +}_{k-1} +\rangle,
\end{equation*}
then $k-1$ contiguous GTP$^+$ monomers next to the tip convert, and finally
the GTP$^+$ at the tip converts to GDP$^-$ before the next attachment event.
The probability for the first $k-1$ conversion events is $\lambda^{-(k-1)}
(k-1)!$, where the factorial arises because the order of these steps is
irrelevant.  The probability of the last step is $\lambda^{-1}$.  Thus the
relative frequency of avalanches of size $k$ is
\begin{equation}
\label{Ak}
A_k\sim \lambda^{-k}\Gamma(k).
\end{equation}

The result can also be derived by the approach of Sec.~\ref{continuum}.  We
use the fact that the configuration
\begin{equation*}
|+\underbrace{-\cdots -}_{k-1} +\rangle
\end{equation*}
occurs with probability $\prod_{1\leq n\leq k-1} (1-e^{-n/\lambda})$.
Multiplying by the probability that the monomer at the tip converts before
the next attachment event then gives the probability for an avalanche of size
$k$:
\begin{equation}
\label{Ak-gen}
A_k = (1+\lambda)^{-1}\prod_{n=1}^{k-1} (1-e^{-n/\lambda})
\end{equation}
Using $1-e^{-n/\lambda}=n/\lambda$ we recover \eqref{Ak}.  If we expand the
exponent to the next order, $1-e^{-n/\lambda}\approx n/\lambda-n^2/(2\lambda^2)$,
Eq.~\eqref{Ak-gen} becomes
\begin{equation*}
A_k = \lambda^{-k}\,\Gamma(k)\prod_{n=1}^{k-1} \left(1-\frac{n}{2\lambda}\right)\sim
\lambda^{-k}\,\Gamma(k)\,e^{-k^2/4\lambda}.
\end{equation*}

\section{GENERAL GROWTH CONDITIONS}
\label{general}

The general situation where the attachment and detachment rates, $\lambda$
and $\mu$, have arbitrary values, and where the parameter $p\ne 1$ seems
analytically intractable because the master equations for basic observables
are coupled to an infinite hierarchy of equations to higher-order variables.
For example, the quantity $n_0\equiv {\rm Prob}\{-\rangle\}$, the probability
for a cap of length zero, satisfies the exact equation
\begin{equation}
\label{n0-mu}
\dot n_0 = -p\lambda n_0+(1-n_0)-\mu\mathcal{N}_0 ~,
\end{equation}
and the speed of the tip is
\begin{equation}
\label{speed-gen}
V(\lambda, \nu, p) = p\lambda n_0 + \lambda(1-n_0)- \mu n_0 ~.
\end{equation}
Here $\mathcal{N}_0\equiv {\rm Prob}\{+-\rangle\}$ is the probability that
there is a GDP$^-$ at the front position with a GTP$^+$ on its left.  Thus to
determine $n_0$ we must find $\mathcal{N}_0$, which then requires
higher-order correlation functions, {\it etc}.  This hierarchical nature
prevents an exact analysis and we turn to approximate approaches to map out
the behavior in different regions of the parameter space.

\subsection{Limiting Cases}

For $\lambda, \mu \ll 1$, the conversion GTP$^+\to$ GDP$^-$ at rate one
greatly exceeds the rates $\lambda, p\lambda, \mu$ of the other three basic
processes that govern microtubule dynamics.  Hence we can assume that
conversion is instantaneous.  Consequently, the end of a microtubule consists
of a string of GDP$^-$, $|\cdots ---\rangle$, in which the tip advances at
rate $p\lambda$ and retreats at rate $\mu$.  Thus from \eqref{speed-gen} the
speed of the tip is
\begin{equation}
\label{V-small}
  V(p,\lambda, \mu) = p\lambda-\mu
\end{equation}
when $p\lambda>\mu$. 
 
On the other hand, for $\lambda\gg 1$, $n_0\equiv {\rm Prob}\{-\rangle\}$ is small
and ${\rm Prob}\{--\rangle\}$ is exceedingly small.  Hence $n_0={\rm
  Prob}\{--\rangle\}+\mathcal{N}_0\approx\mathcal{N}_0$.  Substituting this result
into \eqref{n0-mu} and solving for $n_0$ we find
\begin{equation}
\label{n0-solution}
n_0 =  \frac{1}{1+p\lambda+\mu}.
\end{equation}
Note that indeed $n_0\ll 1$ when $\lambda\gg 1$.  Using \eqref{n0-solution}
in \eqref{speed-gen} we obtain the general result for the growth velocity
\begin{equation}
\label{V-large}
V=\lambda-\frac{(1-p)\lambda+\mu}{1+p\lambda+\mu} 
\quad{\rm when}\quad \lambda\gg 1.
\end{equation}

\subsection{The Phase Boundary}

\begin{figure}[ht]
\includegraphics*[width=0.315\textwidth]{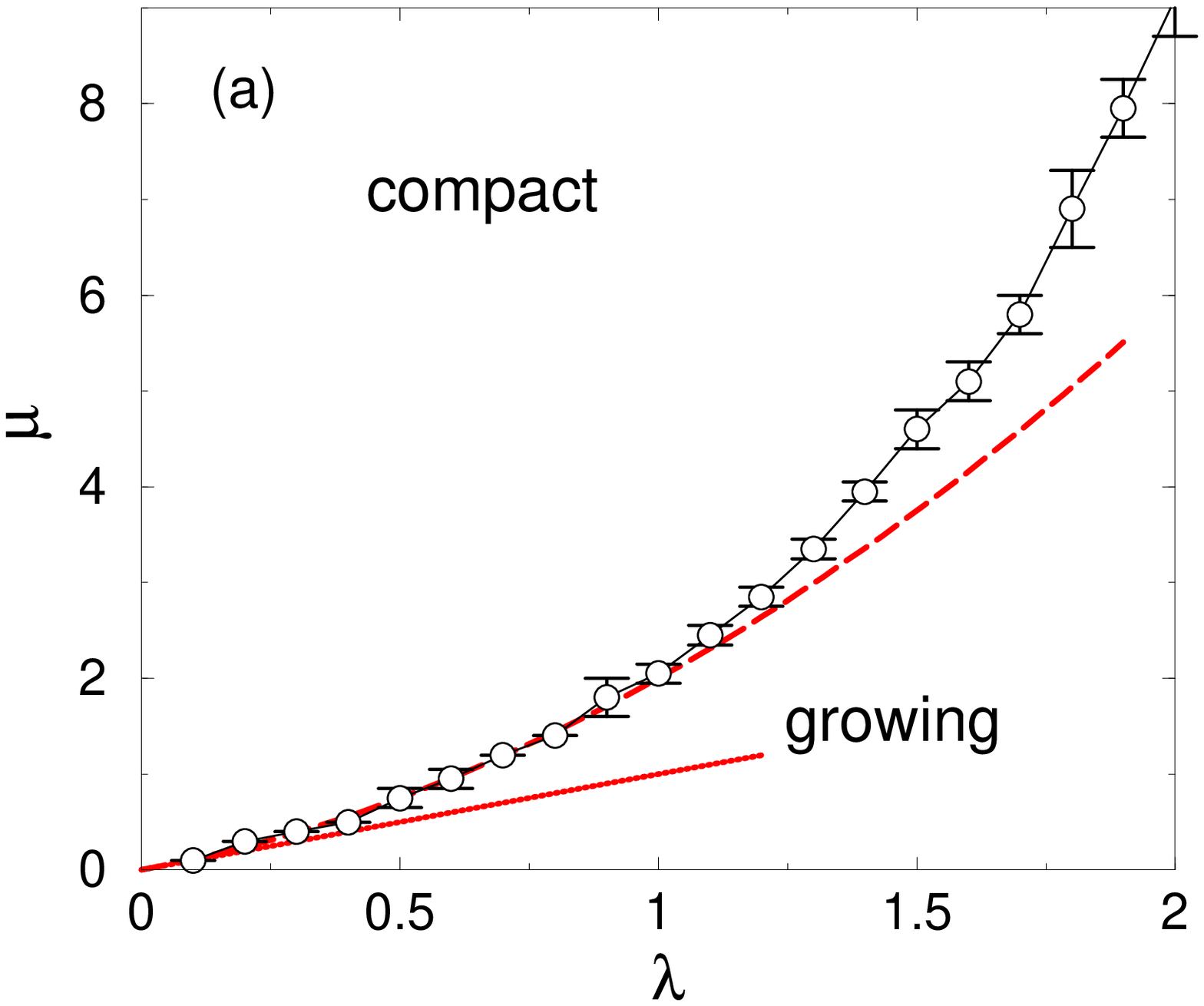}
\includegraphics*[width=0.315\textwidth]{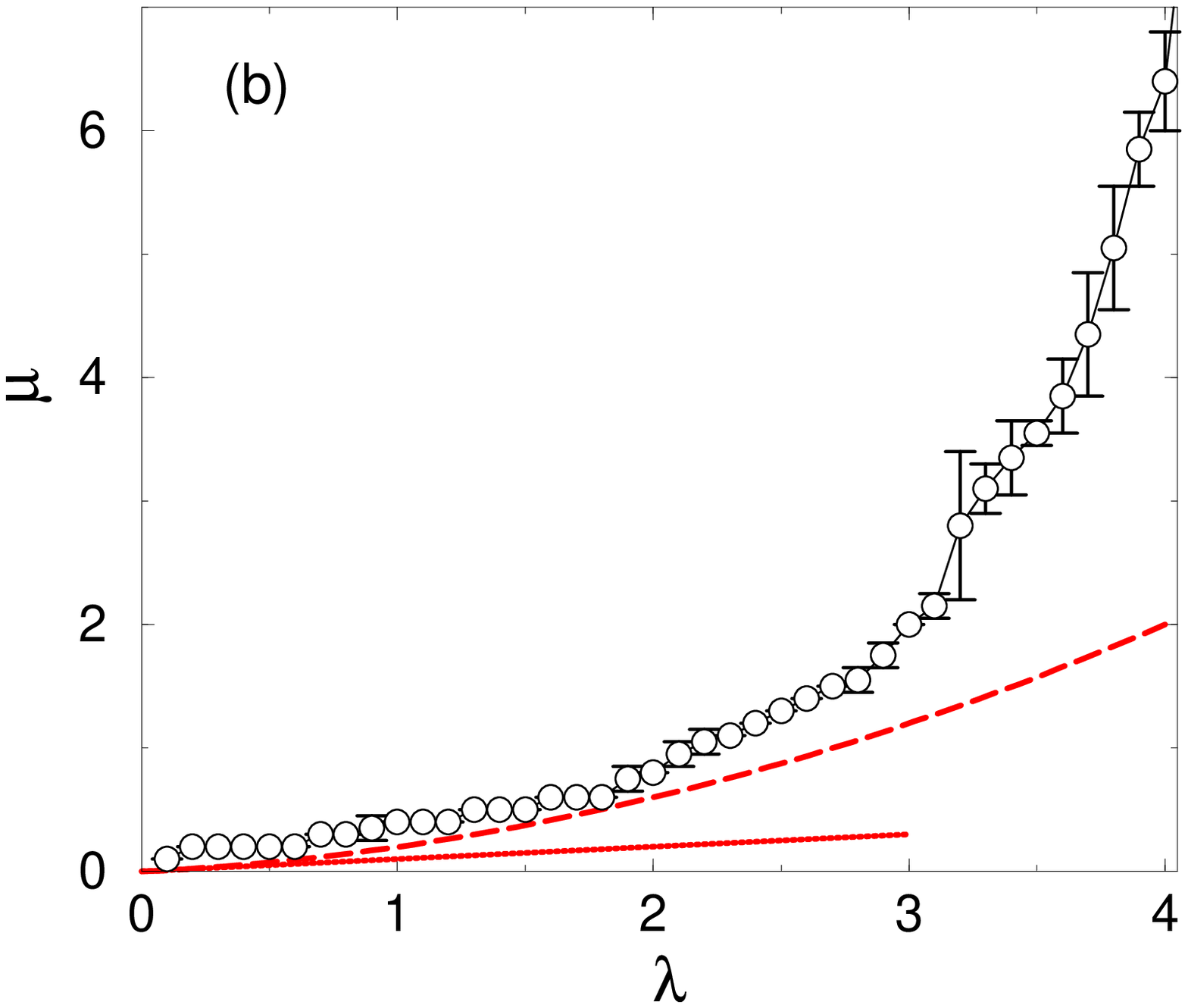}
\caption{Phase diagrams of a microtubule from simulations for (a) $p=1$ and
  (b) $p=0.1$.  The dashed line represents the prediction \eqref{crit-small}
  that is appropriate for small $\mu$.  The extremes of the error bars are
  the points for which the tubule velocities are 0.005 and
  0.015, and their average defines the data point. }
\label{phase-diag}
\end{figure}

A basic characteristic of microtubule dynamics is the phase boundary in the
parameter space that separates the region where the microtubule grows without
bound and a region where the mean microtubule length remains finite.  For
small $\lambda$, this boundary is found from setting $V=0$ in
Eq.~\eqref{V-small} to give $\mu_* = p\lambda$ for $\lambda\ll 1$.  The phase
boundary is a straight line in this limit, but for larger $\lambda$ the
boundary is a convex function of $\lambda$ (see Fig.~\ref{phase-diag}).  We
can compute the velocity to second order in $\mu$ and $\lambda$ by assuming
$\mathcal{N}_0=0$ and then using \eqref{n0-mu} and \eqref{speed-gen}. This
leads to the  phase boundary
\begin{equation}
\label{crit-small}
\mu_*=p\lambda+p\lambda^2,
\end{equation}
that is both convex and more precise.  On this phase boundary, the average
tubule length grows as $\sqrt{t}$.

When $\lambda$ is large, Eq.~\eqref{V-large} implies that $V$ is positive and
it reduces to $V=\lambda-1$ for $\mu\gg \lambda$.  This simple result follows
from the fact that recession of the microtubule is controlled by the unit
conversion rate.  As soon as conversion occurs, detachment occurs immediately
for $\mu\gg\lambda$ and the microtubule recedes by one step.  Since advancement
occurs at rate $\lambda$, the speed of the tip is simply $\lambda-1$.
However, for extremely large $\mu$ the prediction $V=\lambda-1$ breaks
down and the microtubule becomes compact.  To determine the phase boundary in
this limit, consider first the case $\mu=\infty$.  As shown in the previous
section, the probability of a catastrophe roughly scales as
$e^{-\pi^2\lambda/6}$ so that the typical time between catastrophes is
$e^{\pi^2\lambda/6}$.  Since $V=\lambda-1$, the typical length of a
microtubule just before a catastrophe is $(\lambda-1) e^{\pi^2\lambda/6}$.
Suppose now the detachment rate $\mu$ is very large but finite.  The
microtubule is compact if the time to shrink a microtubule of length $\lambda
e^{\pi^2\lambda/6}/\mu$ is smaller than the time $(p\lambda)^{-1}$ required
to generate a GTP$^+$ by the attachment $|\cdots -\rangle \Longrightarrow
|\cdots -+\rangle$ and thereby stop the shrinking.  We estimate the location
of the phase boundary by equating the two times to give
\begin{equation}
\label{crit-large}
\mu_*\sim p \lambda^2 e^{\pi^2\lambda/6}\quad{\rm when}\quad \lambda\gg 1.
\end{equation}

We checked the theoretical predictions \eqref{crit-small} and
\eqref{crit-large} for the phase boundary numerically
(Fig.~\ref{phase-diag}).  For small $\lambda$, the agreement between theory
and the simulation is excellent.  For larger $\lambda$, the tubule growth is
more intermittent and it becomes increasingly difficult to determine the
phase boundary with precision.  Nevertheless, the qualitative expectations of
our theory remain valid.

\subsection{Fluctuations of the Tip}

Finally, we study the fluctuations of the tip in the small and large
$\lambda$ limits.  In the former case but also on the growth phase
$p\lambda>\mu$, the tip undergoes a biased random walk with diffusion
coefficient
\begin{equation}
\label{D-small}
  D(p,\lambda, \mu) = \frac{p\lambda+\mu}{2} 
  \quad{\rm when}\quad 1\gg p\lambda>\mu
\end{equation}

For large $\lambda$, the analysis is simplified by the principle that can be
summarized by: ``The leading behaviors in the $\lambda\to\infty$ limit are
universal, that is, independent of $p$ and $\mu$.'' This is not true if $p$
is particularly small [like $\lambda^{-1}$] and/or if $\mu$ is particularly
large [like $\mu_*$ given by \eqref{crit-large}]. But when $p\lambda\gg 1$
and $\mu\ll \mu_*$, the above principle is true, and
\begin{equation}
\label{VD-sol}
V=\lambda, \quad D=\frac{\lambda}{2}
\end{equation}
in the leading order. 

The computation of sub-leading behaviors is more challenging. We merely state
here two asymptotic results. When $\mu\ll\lambda$, we again have the relation
$D=V/2$, with $V=\lambda+1-p^{-1}$. If $\mu_*\gg \mu\gg\lambda\gg 1$, we
find
\begin{equation}
\label{VD-large}
V=\lambda-1, \quad D=\frac{\lambda+1}{2}
\end{equation}
The derivation of the latter uses the probabilities $X(L,t)$ and $Y(L,t)$ and
follows similar steps as in Sect.~\ref{XY-sub}.

\section{SUMMARY}

We investigated a simple dynamical model of a microtubule that grows by
attachment of a GTP$^+$ to its end at rate $\lambda$, irreversible conversion
of any GTP$^+$ to GDP$^-$ at rate 1, and detachment of a GDP$^-$ from the end
of a microtubule at rate $\mu$.  Remarkably, these simple update rules for a
one-dimensional system lead to steady growth, wild fluctuations, or a steady
state.  Our model has a minimalist formulation and therefore is not meant to
account for all of the microscopic details of microtubule dynamics.  Rather,
our main goal has been to solve for the structural and dynamical properties
of this idealized microtubule model.  Some of the quantities that we
determined, such as island size distributions, have not been studied
previously.  Thus our predictions about the cap and island size distributions
may help motivate experimental studies of these features of microtubules.

A rich phenomenology was found as a function of the three fundamental rates
in our model.  When GTP$^+$ attachment is dominant ($\lambda\gg 1$) and the
attachment is independent of the identity of the last monomer on the free end
of the microtubule ($p=1$), the GTP$^+$ and GDP$^-$ organize into alternating
domains, with gradually longer GTP$^+$ domains and gradually shorter GDP$^-$
domains toward the tip of the microtubule (Fig.~\ref{cartoon}).  Here, the
parameter $\lambda$ could naturally be varied experimentally by either
changing the temperature or the concentration of tubulins (free GTP$^+$) in
the solution.  

The basic geometrical features in this growing phase of a microtubule can be
summarized as:

\medskip
\begin{tabular}{|c|l|c|}
	\hline
symbol &  meaning   &   scaling behavior  \\
	\hline
$N$  & \# GTP$^+$ monomers & $\lambda$  \\
$L$  & tubule length & $\lambda t$  \\
$\langle k\rangle$ & average cap length &  $\sqrt{\pi\lambda/2}$\\
$I$ & \# islands  & $ \lambda/2$\\
$I_k$ & \# GTP$^+$ $k$-islands  & $2\lambda/k^3$\\
$J_k$ & \# GDP$^-$ $k$-islands  & $\lambda/k^2$\\
\hline
\end{tabular}

\smallskip
\noindent We emphasize that the island size distributions of GTP$^+$ and
GDP$^-$ are robust power laws with respective exponents of $3$ and $2$.  In
the limit of $p\ll 1$, in which attachment is suppressed when a GDP$^-$ is at
the free end of the microtubule, the average number of GTP$^+$ monomers on
the microtubule asymptotically is $p\lambda$, while the rest of the results
in the above table remain robust in the long-time limit.

Conversely, when detachment of GDP$^-$ from the end of the tubule is dominant
(detachment rate $\mu\to\infty$, a rate that also could be controlled by the
temperature), the microtubule length remains bounded but its length can
fluctuate strongly.  When the attachment rate is also large, the strong
competition between attachment and detachment leads to wild fluctuations in
the microtubule length even with steady external conditions.  We developed a
probabilistic approach that shows that the time between catastrophes, where
the microtubule shrinks to zero length, scales exponentially with the
attachment rate $\lambda$.  Thus a microtubule can grow essentially freely
for a very long time before undergoing a catastrophe.

For the more biologically relevant case of intermediate parameter values, we
extended our theoretical approaches to determine basic properties of the
tubule, such as its rate of growth, fluctuations of the tip around this mean
growth behavior, and the phase diagram in the ($\lambda,\mu$) parameter
space.  In this intermediate regime, numerical simulations provide more
detailed picture of the geometrical structure and time history of a
microtubule.

\acknowledgments{Paul Weinger generated the early numerical results that led
  to this work.  We also thank Rajesh Ravindran, Allison Ferguson and Daniel
  Needleman for many helpful conversations.  We acknowledge financial support
  to the Program for Evolutionary Dynamics at Harvard University by Jeffrey
  Epstein and NIH grant R01GM078986 (TA), NSF grant DMR0403997 (BC and MM),
  and NSF grants CHE0532969 (PLK) and DMR0535503 (SR).}

\appendix

\section{JOINT DISTRIBUTION FOR $p=1$}
\label{app-P}

Introducing the two-variable generating function 
\begin{equation}
\label{gen-P}
\mathcal{P}(x,y,t)=\sum_{L\geq N\geq 0}x^L y^N P(L,N,t)
\end{equation}
we recast \eqref{PLN} into a partial differential equation 
\begin{equation}
\label{P-diff}
\frac{\partial \mathcal{P}}{\partial t}=(1-y)\frac{\partial \mathcal{P}}{\partial y}
-\lambda(1-xy)\mathcal{P}
\end{equation}
Writing
\begin{equation}
\label{PQ}
\mathcal{P}(x,y,t)=e^{\lambda[xy-(1-x)\ln(1-y)]}\,\,\mathcal{Q}(x,y,t)
\end{equation}
we transform \eqref{P-diff} into a wave equation for an auxiliary function 
$\mathcal{Q}(x,y,t)$
\begin{equation}
\label{Q-diff}
\frac{\partial \mathcal{Q}}{\partial t}=(1-y)\frac{\partial \mathcal{Q}}{\partial y}
\end{equation}
whose general solution is 
\begin{equation}
\label{Q-sol}
\mathcal{Q}(x,y,t)=\Phi(x,\ln(1-y)-t)
\end{equation}
The initial condition $P(L,N,0)=\delta_{L,0}\,\delta_{N,0}$ implies
$\mathcal{P}(x,y,0)=1$ and therefore
\begin{equation*}
e^{\lambda[xy-(1-x)\ln(1-y)]}\,\,\Phi(x,\ln(1-y))=1
\end{equation*}
from which
\begin{equation}
\label{Phi-sol}
\Phi(a,b)=e^{\lambda[(1-a)b+a(e^b-1)]}
\end{equation}
Combining Eqs.~\eqref{PQ}, \eqref{Q-sol}--\eqref{Phi-sol} we arrive at
\begin{equation}
\label{P-sol}
\lambda^{-1}\ln\mathcal{P}=xy(1-e^{-t})-t-x(1-e^{-t}-t)
\end{equation}

\section{JOINT DISTRIBUTION FOR $p\ne1$}
\label{app}

For $p\ne 1$, we consider the distributions $X_N$ and $Y_N$, defined as the
probabilities to have $N$ GTP$^+$ monomers with the tip being either GTP$^+$
or GDP$^-$, respectively.  These probabilities satisfy a closed set of
coupled equations.  In the stationary state these equations become
\begin{subequations}
\begin{align}
\label{XN}
   &(N+\lambda) X_N=\lambda X_{N-1}+p\lambda Y_{N-1}+NX_{N+1}\\
\label{YN} 
   &(N+p\lambda) Y_N=X_{N+1}+(N+1)Y_{N+1}.
\end{align} 
\end{subequations}

Since $X_0\equiv 0$, it is convenient to define the generating functions
corresponding to $X_N$ and $Y_N$ as follows:
\begin{subequations}
\begin{align}
\label{X-gen-def}
\mathcal{X}(z)&= \sum_{N\geq 1}z^{N-1} X_N\\
\label{Y-gen-def}
\mathcal{Y}(z)&=\sum_{N\geq 0}z^{N} Y_N.
\end{align} 
\end{subequations}
Now multiply Eq.~\eqref{XN} by $z^{N}$ and Eq.~\eqref{YN} by $z^{N-1}$ and
sum over all $N\geq 1$ or $N\geq 0$, respectively, to obtain
\begin{subequations}
\begin{align}
\label{XY1}
   p\lambda \mathcal{Y}&=\mathcal{X}-\zeta(\mathcal{X}-\mathcal{X}')\\
\label{XY2} 
   p\lambda \mathcal{Y}&=\mathcal{X}-\zeta\mathcal{Y}'.
\end{align} 
\end{subequations}
where $\zeta=\lambda(z-1)$ and prime denotes a derivative in $\zeta$.  We can
reduce these two coupled first-order differential equations to uncoupled
second-order equations:
\begin{subequations}
\begin{align}
\label{X-gen}
   &\zeta\mathcal{X}'' + (2+p\lambda-\zeta)\mathcal{X}'-(1+p\lambda)\mathcal{X}=0\\
\label{Y-gen} 
   &\zeta\mathcal{Y}'' + (2+p\lambda-\zeta)\mathcal{Y}'-p\lambda\mathcal{Y}=0.
\end{align} 
\end{subequations}
The solutions are the confluent hypergeometric functions
\begin{subequations}
\begin{align}
\label{X-gen-sol}
   \mathcal{X}(z)&=\frac{p\lambda}{1+p\lambda}\,F(1+p\lambda; 2+p\lambda; \zeta)\\
\label{Y-gen-c} 
    \mathcal{Y}(z)&=\frac{1}{1+p\lambda}\,F(p\lambda; 2+p\lambda; \zeta).
\end{align} 
\end{subequations}

These generating functions have seemingly compact expressions but one has to
keep in mind that the $X$ and $Y$ probabilities are actually infinite sums.
For instance, $Y_0=\mathcal{Y}(z=0)=\mathcal{Y}(\zeta=-\lambda)$. Recalling
the definition of the confluent hypergeometric function we obtain
\begin{eqnarray*}
 Y_0 &=&\frac{1}{1+p\lambda}\,F(p\lambda; 2+p\lambda; -\lambda)\\
 &=& \frac{1}{1+p\lambda}\sum_{n\geq 0}
 \frac{(p\lambda)_n}{(2+p\lambda)_n}\,\frac{(-\lambda)^n}{n!}
\end{eqnarray*}
where $(a)_n=a(a+1)\ldots (a+n-1)=\Gamma(a+n)/\Gamma(a)$ is the rising factorial. 
Note that $\Pi_0=Y_0$. Computing 
\begin{eqnarray*}
X_1 &=&\frac{p\lambda}{1+p\lambda}\,F(1+p\lambda; 2+p\lambda; -\lambda)\\
Y_1 &=&\lambda\,\frac{p\lambda}{(1+p\lambda)(2+p\lambda)}\,
F(1+p\lambda; 3+p\lambda; -\lambda)
\end{eqnarray*}
one can determine $\Pi_1=X_1+Y_1$.  Some of these formulas can be simplified
using the Kummer relation
\begin{equation*}
F(a;b;\zeta)=e^\zeta F(b-a;b;-\zeta)
\end{equation*}
For instance,
\begin{eqnarray*}
  Y_0 &=&\sum_{n\geq 0}
  \frac{(n+1)\lambda^n e^{-\lambda}}{(1+p\lambda)_{n+1}}\\
  X_1 &=&p\lambda\sum_{n\geq 0}
  \frac{\lambda^n e^{-\lambda}}{(1+p\lambda)_{n+1}}\\
  Y_1 &=&p\lambda^2
  \sum_{n\geq 0}\frac{(n+1)\lambda^n e^{-\lambda}}{(1+p\lambda)_{n+2}}.
\end{eqnarray*}


\begin{thebibliography}{19}
\bibitem{dimer} We term a heterodimer that consists of $\alpha$- and
  $\beta$-tubulin as a ``monomer'' throughout this paper.
\end{thebibliography}
\end{document}